# GLOBAL PUBLIC GOODS: THE CASE FOR THE GLOBAL EARTH OBSERVATION SYSTEM OF SYSTEMS
___________________________________________________________________

**Miloslav Machoň**
___________________________________________________________________

**Abstract**
The debate surrounding the provision of welfare by state institutions has been widely discussed in the field of political economics since the 1930s. Related research also focuses on welfare supply at an international system level. This article assesses whether international cooperation in the area of sharing remote sensing data leads to the supply of global public goods, which to date has not yet been discussed in related scholarly literature. The supply of global public goods is assessed within the GEO international regime and leads to the use of the non-rivalrous GEOSS, which can be accessed by every socio-economic group in every UN member country including future generations. However, providing the benefit of GEOSS is not always favourable because of the low number of financially participating consumers.

## Introduction

Following the Great Depression, political economy concentrated on the question of the state's role in economy [Pigou 1932; Keynes 1936; Musgrave 1939]. This subject gained relevance after the Second World War when Keynesian macroeconomics [Robinson − Eatwell, 1973; Eichner − Kregel, 1975] became a prevailing paradigm. It emphasised, along with the welfare economics [Sen, 1963; Graaff, 1968; Harberger, 1971], an active role of the state in planning and controlling the economy. In the late 20th century political economy thus developed concepts which revised the models of classical macroeconomics about the expectations and uncertainty in monetary economics [Shackle 1970] or about factors affecting the work productivity [Schapiro − Stiglitz, 1984], as well as classical economics models about the rational nature and expectations of an individual [Mankiw − Romer, 1991].

Prioritising an active role of the state in planning and controlling the economy also renewed the political economy debate about the state's responsibility to support the economic welfare of individual national entities [Lindahl, 1958; Mazzola, 1958; Sax, 1958]. This debate followed the Austrian-German and Italian traditions of political economy, which at the turn of the 20th century focused on the distribution of state expenditures [Schulze-Gävernitz, 1899; Brentano, 1910]. The issue of the distribution of national expenditures was researched by Richard Musgrave, who defined the welfare provided by national entities as a collective service or "good" [Musgrave, 1939: 215].

Paul Samuelson [1954: 387] used Musgrave's concept of a good when he drew attention to collective goods as opposed to private goods. Samuel's model of collective goods using Musgrave's model of social needs [1959: 10] was further developed by Head [1962: 199-207] when he created a model of public goods whose consumption is indivisible, and the consumers cannot be excluded from it. Following the processes of globalisation in the late 1990s, Head's model was extended further [Kaul et al., 2002: 12] as the benefits of national public goods began to spill beyond a territory of a country. New categories of public goods, including local and global public goods, developed in political economy based on the spatial extent of their benefits [Kaul et al., 1999a: 10-11; Kaul − Mendoza 2002: 89-94; Morrissey et al., 2002: 36-37].

A number of existing studies work with the model of global public goods. The studies focus on individual areas of globalisation, especially on the issue of international healthcare provision as a global public good [Chen et al., 1999; Smith et al., 2003; Brown et al., 2006; Moon, 2009; Weir −Mykhalovskiy, 2014], on the provision of global public goods by a civil society [Einarsson, 2001; Kaul, 2001a; Henderson, 2002], on the role of global public goods in providing development aid [Raffer, 1999; Kaul, 2001b; Morrissey et al., 2002; Guillaumont, 2002; Kaul 2005; Mascarenhas − Sandler, 2005], on the status of natural environment as a global public good [Nordhaus, 1999; Saurin, 2001; Unnevehr, 2004; Milinski et al., 2006], or on the status of energy security as a global public good [Gupta − Ivanova, 2009; Goldthau, 2011; Karlsson-Vinkhuyzen et al., 2012]. Many areas of globalisation, however, have not been so far analysed using the theory of global public goods.

One of these areas is international cooperation in sharing of the data of the remote Earth sensing research. The research is carried out from the outer space, which is a global public good [Vogler, 2012: 172-176]. We can thus assume that international cooperation in the field of using satellite technologies also aims to create global public goods, although it has not yet been confirmed by any of the existing research. The objective of this paper is thus to examine whether international cooperation in sharing remote sensing data results in a provision of a pure global public good. This objective is achieved by using the analysis of basic aspects of providing benefits within the Global Earth Observation System of Systems (GEOSS). We examine to what extent the provision of the GEOSS constitutes a benefit of a pure global public good from the perspective of its spatial extent and from the perspective of meeting the criteria of indivisibility of its consumption and the impossibility of excluding the consumer from its consumption. We also focus on the question whether the GEOSS is provided in an optimum amount.

In order to meet the stated objective, the paper progresses in three steps. The first part focuses on the characteristics of the public goods and analyses their influence on the problem of market failure in the provision of public goods [see Samuelson 1954; 1955; 1969; Head 1962; 1977]. The second part constitutes a theoretical framework for identification of public goods and their provision on the global level [Kaul et al., 1999a; Kaul − Mendoza, 2002; Sandler, 2002b]. The third part examines to what extent the benefits of the GEOSS fit the characteristics of the benefits of a global public good. In order to do so, it uses the constitutive documents of the Group on Earth Observations (GEO), framework documents of the GEOSS, and reports on its funding.

Due to its limited scope, the paper only analyses the sharing of remote sensing data on the international level and not on the level of individual UN member states. The analysis does not take into account e.g., the level of access to IT technologies or qualified human resources in the UN member countries that are necessary for the access to the benefit of the GEOSS. It cannot unequivocally demonstrate a causal relationship between the use of remote sensing data and its socio-economic impact because based on the available sources it is impossible to exclude the contribution of other factors affecting the level of the socio-economic impact resulting from the access to the remote sensing data (the economic profile of the country, influence of economic cycles, influence of the process of global warming etc.).

# 1 Provision of public goods and the spatial extent of their benefits

Concept of a public good was created as a synthesis and expansion of the concepts collective good and social need [Head, 1962: 202]. Collective good [Samuelson, 1954: 387; 1958: 333-335] is the opposite of private good based on the divisibility of consumption. While the consumption of a private good is divided between individual consumers, the consumption of a collective good is collective. Its consumption by one consumer does not decrease the benefits available to other consumers. A social need [Musgrave, 1959: 9-10] is also met collectively. The concept of a social need is based on the perspective of a collective of consumers. It captures more accurately the relationship between the contribution of an individual consumer and the benefit created by meeting social needs. Its amount does not depend on an individual contribution of a consumer to the meeting of the social needs, but on the amount of welfare of a community of consumers [Musgrave, 1959: 9-12].

A global public good is characterised by the fact that (1) its consumption by one consumer does not decrease the available benefit for other consumers and (2) it is not possible to exclude a consumer from the consumption of its benefits [Head, 1962: 204-205]. The indivisibility of consumption and the impossibility to exclude a consumer from the consumption of the good are mutually independent characteristics with equal relevance for the identification of a public good [Head, 1962: 208]. In the provision of a public good, the sum of marginal rates of substitution (MRS) corresponds to the marginal rate of transformation: (MT) $\sum \llbracket MRS=MT \rrbracket$ [Samuelson, 1954: 387-388; 1955: 353-354]. This is different from the provision of private goods, where the marginal rate of transformation corresponds to an individual value of the marginal rate of substitution: $MRS_1 = MRS_2 = MT$ [Samuelson, 1954: 354]. The market cannot thus contribute to the provision of the public good, only a public central authority superior to the market can [Samuelson, 1954: 377-388; 1958: 334-337; 1969: 28-29; Musgrave, 1959: 9-10]. This especially affects the impossibility to exclude an individual from the consumption of the good's benefits [Head, 1977: 234-235].

If the impossibility to exclude an individual from the consumption of the good's indivisible benefit is spatially determined by the state borders, then it is a national public good (see Table 1) [Morrissey et al., 2002: 36-37; Kaul − Mendoza, 2002: 94, 107; Stiglitz, 2013: 253; Frieden, 2016: 3,6]. Its consumers are socio-economic segments of the country's population and the provider is a central authority on the national level which is superior to the market. Due to globalisation, however, the benefit of the national public good spatially "spills over" the state borders [Kaul et al., 2002: 12]. If the good is consumed also by socio-economic segments of populations and governments of neighbouring countries, it is then a regional public good [Morrissey et al., 2002: 34]. A public good whose benefit extends beyond the territory of a regional public good is a global public good [Barrett, 2002: 52-53].

The provider of a global public good is not usually a central authority on the national level (as it is in the case of national and regional public goods), but a central authority on the level of the international system superior to the market – an international regime [Krasner, 1982: 186; Baert et al., 2014: 5-6] allowing joint participation of countries and international governmental organisations in international cooperation. Its shape is determined by a set of rules (i.e., abstract objects defining the manner of behaviour). [Boghossian, 2015: 4], norms (i.e., abstract objects constituting acceptable or forbidden behaviour) [Hage, 2015: 14], principles (i.e., abstract objects constituting acceptable or forbidden behaviour using a moral reason) [Bix, 2015: 135-142], and decision-making procedures (i.e., negotiations leading to creation and implementation of political decisions) [Ciot, 2014: 64-65; Persson, 2015: 2; Pulkowski, 2014: 82]. Public goods are compared in Table 1.

**Table 1 | Types of public goods according to the spatial extent of their benefit**

|  | **Provider** | **Consumer** |
|---|---|---|
| **National public good** | Central authority on the national level | Socio-economic segments of the country's population |
| **Regional public good** |  | Governments and socio-economic segments of the populations of neighbouring countries |
| **Global public good** | Central authority on the international level (i.e., an international regime) | Governments, socio-economic segments of the population, and future generations of the 193 UN member states |

## 2  Global public goods

The following groups have equal access to the benefit of a global public good: (1) governments of all the countries in the world, i.e. governments of the 193 UN member states [UN 2015], (2) socio-economic groups in individual countries, i.e. population segments in the 193 UN member states [Kaul et al., 1999a: 10-11], (3) all generations in the individual countries [Brundtland, 1987: 16-17; Kaul − Mendoza, 2002: 89-94; Louafi, 2015: 85]. Global public good is provided in the way that does not limit the extent of its benefit for future generations of the 193 UN member countries.

If all these criteria are met by the global good, then the good is provided in its pure form [Sandler, 2002b: 85-90]. A consumer contributes to its funding by their marginal willingness to pay [Sandler, 2002b: 83-84; Baumgärtner et al., 2015: 8-10]. A consumer's decision about the specific amount of the contribution is comparable to a prisoner's dilemma. It depends on the individual expenses contributed and individual benefits of the consumer stemming from the consumption of a pure global public good [Sandler, 1992: 36-38; Pinheiro et al., 2014: 2]. The consumer does not consider the total sum of benefits that the consumption of a pure global public good will bring to all consumers. The individual consumers' marginal willingness to pay accumulate. Each individual marginal willingness to pay proportionally affects the total amount of benefit of pure global public good [Sandler, 2002b: 95-97; Sonntag, 2014: 143]. If the sum total of individual consumers' marginal willingness to pay corresponds to the marginal costs of providing a pure global public good, the global public good is then provided in an optimum amount [Sandler, 2002b: 84; Tavor − Spiegel, 2016: 119].

The non-optimum benefit of the provision of a global public good may result from its characteristics, the relationship between the marginal value of the benefit of a public good for individual consumers and the sum of its benefits, or the provider and consumer strategies [Kaul et al., 1999b: 464-468; Sandler, 2002b: 82-85]. Non-excludability from the consumption of a pure global public good causes the consumer's reluctance to disclose the real value of their benefits resulting from the consumption, which affects the marginal willingness to pay [Sandler, 2002b: 83]. The marginal costs of providing pure global public good determined by the provider will thus be lower than marginal costs of providing the good in the optimum amount. [Sandler, 2002b: 84]. This is also caused by the fact that an individual marginal willingness to pay proportionally increases the total amount of benefits of a pure global public good because it is substituted with individual marginal willingness to pay of other consumers [Sandler, 2002a: 134-135; Sonntag, 2014: 143-144]. When a consumer makes a decision about a specific amount of a marginal willingness to pay, this contributes to the choice of the free-rider strategy [Sandler, 1992: 38-44; Ersoy, 2011: 228]. In this way, the consumer will decrease the sum of the benefits of a pure global public good. The provider will cause a non-optimum provision of the benefit of a pure global public good by underprovision [Conceição, 2002: 155] if they deliberately provide its benefit in a decreased or zero amount, if they distort the benefit to the detriment of the consumer, or if they provide the benefit in a way that leads to its overuse [Kaul et al., 1999: 467; Shepsle − Weingast, 2014: 6-7]. If the consumer does not have access to the benefit of a pure global public good due to their financial and technological capabilities or formal restrictions by the provider [De Lombaerde − Langenhove, 2011: 112-113], it will result in underuse [Conceição, 2002: 154-155] on the consumer's end.

Summing up, a global public good is a good which meets the following requirements: 1. The provision of its benefit is managed by an international regime [Krasner, 1982; Boghossian, 2015; Hage, 2015; Bix, 2015] 2. The provision of the benefits of the good meets the defining criteria of a pure global public good when it comes to the indivisibility of its consumption and the impossibility to exclude from its consumption the governments, socio-economic segments of the population, and the future generations of the 193 UN member states [Kaul et al., 1999a; Kaul − Mendoza, 2002; Sandler, 2002b]. When these conditions are met, we can examine whether the benefits of the good are provided in their optimum amount [Sandler, 2002b; Tavor − Spiegel, 2016]. If the benefits of the good are provided in a non-optimum amount, we need to identify the possible reasons for this state of things, including the contribution of the characteristics of the public good, the relationship between the individual marginal value of its benefits to the sum total of its benefits, and the provider and consumer strategies [Kaul et al., 1999b; Sandler, 2002b; Conceição, 2002]. We will analyse the GEOSS from this perspective in the following section of this paper.

## 3  The Global Earth Observation System of Systems as a global public good

The GEOSS is a result of international cooperation in the remote sensing. This cooperation constituted one of the three declarations of an informal G8 group which aimed to reinforce sustainable global development [G8, 2003]. At a summit in Evian, France, in June 2003, the G8 representatives agreed that cooperation in creating and gathering reliable remote sensing data with information on the state of the Earth's atmosphere, soil, drinking water, oceans, and planetary ecosystems is necessary for the support of sustainable development. This declaration was then followed with the first Earth Observation Summit [WMO, 2003: 3] in July 2003, where the first inter-governmental Group on Earth Observations (GEO) was created ad hoc. The group developed a 2005-2015 Implementation Plan, which set out the international cooperation in creating and gathering remote sensing data [GEO, 2015].

There had been a total of six ad hoc meetings of the GEO [2005a] before it was formally established as a permanent platform for international cooperation in creating and gathering remote sensing data at the Third Earth Observation Summit [GEO, 2005b: 1]. The GEO is an international regime, in which the shape of the international cooperation depends on the method of complex, coordinated, and sustainable observation of the planetary system [GEO, 2005c: 1]. The cooperation is acceptable if it leads to the improvement of the state of monitoring of the planetary system, if it increases the understanding of the planetary processes, and if it enhances predictions about the dynamics of the planetary system and planetary processes [GEO, 2005c: 1]. Advancement of the state of observation of the planetary system, increasing the understanding of planetary processes, and enhancement of the predictions about their dynamics aim to meet moral objectives: reducing losses of human lives and property due to natural and man-made disasters, learning about the environmental factors affecting the level of people's health and welfare, improvement of management of water and energy sources, improving the of management and protection of terrestrial, coastal, and marine ecosystems leading to preserving biodiversity [GEO, 2005c: 1-2].

The members of the GEO can be the UN member states and the European Commission. Membership of the European Commission, according to the 2005-2015 Implementation

Plan, is equal to the UN membership [GEO, 2005b: 2-3]. It is executed as part of shared competencies in the area of research and innovation between the EU institutions and its membership countries that are at the same time members of the UN [EC, 1992: 8-9; EC, 2007: 49-50; Kaddous, 2015: 8-10]. In the period between 2005 and 2015, the EU was actively represented in the GEO by the European Commission's Directorate-General for Research and Innovation – a supranational executive EU institution for implementation of policies in this sector [Schmidt − Wonka, 2012: 337-338; Reillon, 2015: 9-10]. The status of a GEO participating international organisation can be obtained by a regional international inter-governmental or non-governmental organisation which deals with the subject of remote Earth sensing.

The negotiations leading to the formulation and implementation of a political decision in the GEO take place in a two-level decision-making structure. [GEO, 2015a]. The first level is a plenary [GEO, 2014a: 2-3], which is the main decision-making body. It assembles at least once a year with actors of international relations including GEO members and participating international organisations attending. The second level of the internal decision-making structure of the GEO is represented by the executive committee, the GEO secretariat with the programme board. The executive committee [GEO, 2014a: 4-5] consists of 13 GEO members nominated by the plenary according to a regional key. The committee implements accepted decisions of the plenary in the periods between the plenary assemblies. Similarly to the plenary, the decisions in the committee are approved by a consensus. The executive committee is also responsible for the work of the GEO secretariat [GEO, 2014a: 5-6], which is managed by the director of the GEO. The director, like the members of the executive board, is nominated by the plenary by consensus for a term of maximum three years. The main role of the GEO secretariat is to run the everyday work of the GEO including its outside representation. The last element of the second level of the internal decision-making structure are implementation boards and working groups [GEO, 2014a: 6-7]. They are also appointed by the plenary. Their main role is a supervision of the implementation of the GEO's professional activities. The specific definition of their workload and the length of their term depends on the respective decisions by the plenary.

The role of the international regime GEO is to provide a benefit in the form of the GEOSS products [GEO, 2005c: 1]. These are raw data, metadata, and end results of the remote Earth sensing [GEO, 2005c: 3]. Their consumption is indivisible because according to the copyright law determining the conditions of their use, they are available to be used and reproduced for free [GEO, 2015b]. They can only be used for personal and non-commercial purposes as the copyright law explicitly prohibits a commercial use of the GEOSS products and their derivatives, including re-sale of the products [GEO, 2015b].

The GEOSS benefit constituted by its products including raw data, metadata, and end results of the remote Earth sensing is available to the government policymakers in the UN member states [GEO, 2005c: 3]. It can also be used by individual socio-economic groups in the UN member states, including researchers, engineers, and representatives of a civic society from governmental and non-governmental organisations. The GEOSS benefit is also available to the future generations in the UN member states since an open archival data storage [GEO, 2005c: 8], where the individual elements of its benefit are gathered, is also a part of the GEOSS

The consumers contribute to the funding of the GEOSS benefit by voluntary payments to the GEO-WMO trust fund, according to their marginal willingness to pay [GEO, 2014a: 9-10]. The marginal willingness to pay depends on the socio-economic impact of the GEOSS benefits. The impact is evident in the use of the GEOSS benefit for meteorological modelling because it provides more accurate input parameters for meteorological and climatological forecasts [Hertzfeld − Williamson, 2007: 244-249].

The individual consumers' marginal willingness to pay accumulates and each individual marginal willingness to pay proportionally affects the total amount of the GEOSS benefit. It is evident in the portfolio of direct and indirect financial costs used for the provision of the GEOSS benefit [GEO, 2014a: 9-10]. The direct costs are expenses connected directly to the provision of the GEOSS benefit such as salaries, travel expenses, and equipment for the employees of the GEO secretariat. The indirect costs are the ones paid for the GEO administrative work, such as preparation of internal research, reports, and auxiliary materials.

The difference between the marginal costs of the provision of the GEOSS benefit and the sum total of the individual consumers' marginal willingness to pay results in a non-optimum provision of the GEOSS benefit. It was apparent in the results of a survey on problems in providing of the GEOSS benefit from the consumers' perspective [Heumesser et al., 2012: 247-249]. In the survey, 44% of consumers indicated that the problem of providing the GEOSS was insufficient institutional support, while 23% said that it was insufficient financial support.

The reasons for non-optimum provision of the GEOS benefit are both the impossibility to exclude a consumer from its consumption and the unwillingness of the consumers to reveal the real value of their marginal benefit resulting from this consumption. The impossibility to exclude a consumer from the consumption of the GEOSS benefit is a result of the free access to the raw data, metadata, and end results of the remote Earth sensing via an internet website [GEO, 2015e]. The unwillingness of the consumers to reveal the real value of their marginal benefit resulting from its consumption stems from the voluntary contribution of This is the translated version of the paper that was already published in Czech language. Please Please cite as follows: Machoň, Miloslav (2017). Global Public Goods: The Case for the Global Earth Observation System of Systems. Acta Oeconomica Pragensia, 25(3), 68-83. https://doi.org/10.18267/j.aop.583payments into the GEO-WMO trust fund [GEO, 2014a: 9-10]. It is evident in the annual reports about the state of the GEO-WMO trust fund [GEO, 2006; GEO, 2007; GEO, 2008; GEO, 2009; GEO, 2010; GEO, 2011; GEO, 2012; GEO, 2013; GEO, 2014b; GEO, 2015f], according to which in the period of 2006–2015 only 21 UN member states and the European Commission contributed to the fund (see Figure 2). The European Commission deposited financial contributions in the GEO-WMO trust fund as part of the EU Framework Programme for Research and Innovation in 2006 and the Horizon 2020 programme between 2007 and 2015 [EC, 2014: 9-11; EC, 2015]. The funding of both programmes was part of the implementation of annual EU budgets in the period 2006-2015, which was the responsibility of the European Commission's Directorate-General for Research and Innovation [EC, 2007: 128; EC, 2012: 140; Reillon, 2015: 8-9].

The total sum of contributions by individual consumers differs, ranging from 7 720 CHF (Mexico) to 7 620 628 CHF (the European Commission) (see Figure 1). There are also significant differences in the distribution of the contributions in the period of 2006-2015, ranging from 92 100 CHF (2013) to 4 923 050 CHF (2014) (see Figure 2). A low number of UN member states willing to reveal the value of their marginal benefit resulting from the consumption of the GEOSS benefit along with the differing amounts of individual contributions results in assigning lower expenses for providing the GEOSS benefit than the marginal expenses for its provision in optimum amount. Consumers' individual marginal willingness to pay into the GEO-WMO trust fund proportionally increases the total amount of the GEOSS benefit as it is substituted by the individual marginal willingness to pay of other consumers [GEO, 2014a: 9-10]. This, however, increases the choice of the free-rider strategy, which decreases the total amount of the GEOSS benefit. The free-rider strategy is prevalent among the GEOSS benefit consumers when making a decision about the specific amount of individual marginal willingness to pay because 172 UN member states so far have not contributed to the funding of the GEOSS benefit [GEO, 2006; GEO, 2007; GEO, 2008; GEO, 2009; GEO, 2010; GEO, 2011; GEO, 2012; GEO, 2013; GEO, 2014b; GEO, 2015f].

**Figure 1 | Finances deposited in the GEO-WMO trust fund (CHF) in the period of 2006–2015**

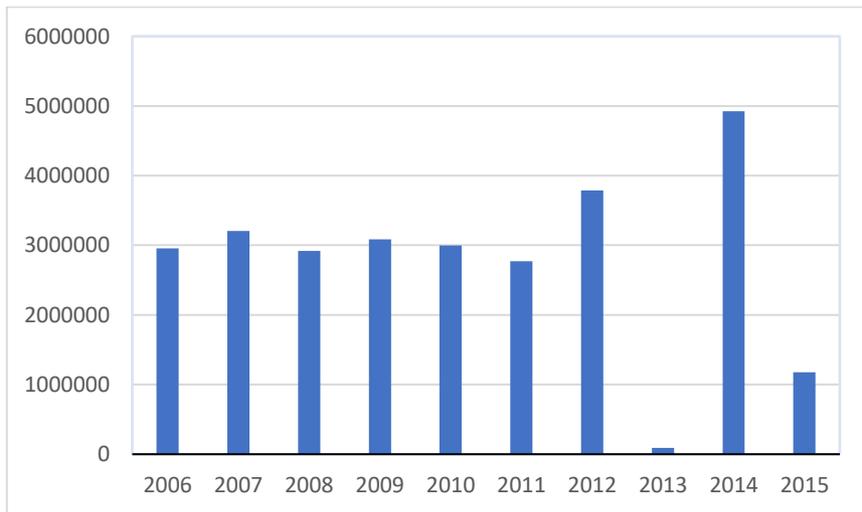

**Figure 2 | Finances deposited in the GEO-WMO trust fund (CHF) in the period of 2006–2015**

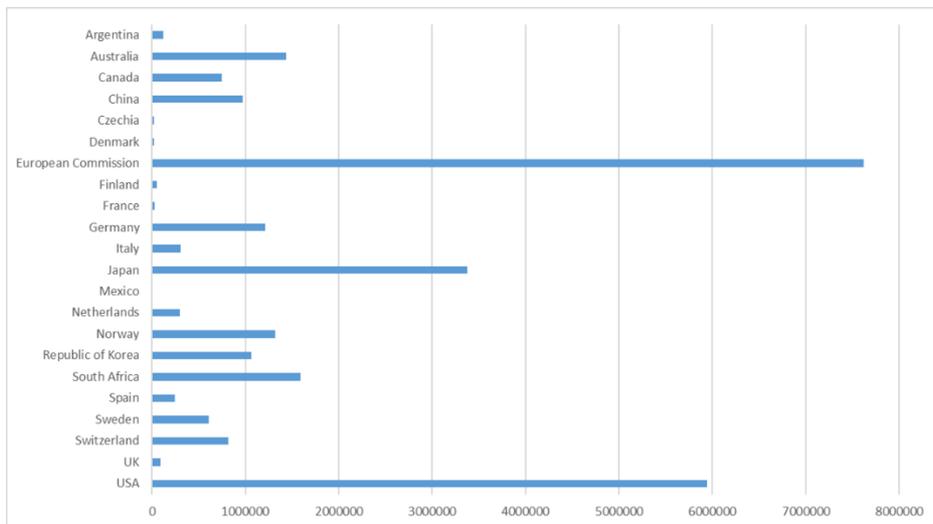

The reason for the non-optimum provision of the GEOSS benefit is not the decreased provision on the part of the provider or decreased consumption. According to the survey on problems in providing of the GEOSS benefit, it is not distorted to the consumer's disadvantage [Heumesser et al., 2012: 247-249]. 81% of consumers perceive the influence of the GEOSS benefit on their work as satisfactory and according to 76% of consumers the GEOSS benefit speeds up the work on joint projects. Furthermore, the access to the GEOSS benefit is not denied to the consumers due to their financial and technological capabilities or formal restrictions on the part of the provider, as 62% of consumers say that the GEOSS benefit is easily accessible and 82% of consumers consider useful the fact that technological convertibleness of the GEOSS benefit useful [Heumesser et al., 2012: 247-249].

## Conclusion

The international cooperation in the area of sharing of remote sensing data aims to create global public goods as the provision of the GEOSS meets these characteristics: (1) Provision of the GEOSS benefit is a role of an international regime GEO, which accepts cooperation leading to improvement of knowledge and refining the forecasts about the planetary processes. (2) The GEOSS benefit is indivisible because it is available for free for non-commercial use via a n internet website. Its spatial extent is global as it can be accessed by governments, socio-economical segments of populations, and the future generations of the 193 UN member states.

The GEOSS benefit is provided in a non-optimum amount. The reasons for this non-optimum provision is the impossibility to exclude a consumer from its consumption, as it is provided for free, and the voluntary funding resulting from a low level of international cooperation. Only a low number of consumers contributes to the funding of the GEOSS benefit. Moreover, the individual marginal will to pay is substituted by the individual marginal will to pay by other consumers. Decreased provision on the part of the provider or decreased consumption are not the reasons for non-optimum provision of the GEOSS benefit.

The limited scope of this analysis does not allow us to assess the actual impact of the sharing of the remote sensing data in the GEO for individual countries. This implies that political economy should examine this issue in detail. Upcoming research should thus focus on creating a relevant framework for assessing the socio-economic impact resulting from the access to the remote sensing data in the GEO, which would then allow to assess the extent of the impact in the UN member states.

This is the translated version of the paper that was already published in Czech language. Please Please cite as follows: Machoň, Miloslav (2017). Global Public Goods: The Case for the Global Earth Observation System of Systems. Acta Oeconomica Pragensia, 25(3), 68-83. https://doi.org/10.18267/j.aop.583# Literature

BAERT, F.; SCARAMAGLI, T.; SÖDERBAUM, F. Introduction: Intersecting Interregionalism. In: BAERT, F.; SCARAMAGLI, T.; SÖDERBAUM, F. (Eds.) *Intersecting Interregionalism: Regions, Global Governance and the EU*. Dordrecht: Springer. 1st ed. 2014, pp. 1-12. ISBN 978-94-007-7566-4. DOI: 10.1007/978-94-007-7566-4.

BARRETT, S. Supplying International Public Goods: How Nations Can Cooperate? In: FERRONI, M.; MODY, A. (Eds.) *International Public Goods. Incentives, Measurement and Financing*. New York: World Bank Publications. 1st ed. 2002, pp. 47-81. ISBN 0-8213-5110-9.

BAUMGÄRTNER, S.; CHENB, W.; HUSSAINA, A. Willingness to pay for public environmental goods under uncertainty. A paper presented at the 21st Annual Conference of the European Association of Environmental and Resource Economists, Helsinki, Finland, 24-27 June 2015, pp. 1-33.

BIX B. Rules and Normativity in Law. In: ARASZKIEWICZ, M.; BANAS, P.; GIZBERT-STUDNICKI, T.; PLESZKA, K. (Eds.) *Problems of normativity, rules and rule-following.* Springer International Publishing, 1st ed. 2015, pp. 125-146. ISBN 978-3-319-09374-1. DOI: 10.1007/978-3-319-09375-8_10.

BOGHOSSIAN, P. Rules, Norms and Principles: A Conceptual Framework. In: ARASZKIEWICZ, M.; BANAS, P.; GIZBERT-STUDNICKI, T.; PLESZKA, K. (Eds.) *Problems of normativity, rules and rule-following.* Springer International Publishing, 1st ed. 2015, pp. 3-11. ISBN 978-3-319-09374-1. DOI:10.1007/978-3-319-09375-8_1.

BRENTANO, L. The doctrine of Malthus and the increase of population during the last decades. *The Economic Journal*. 1910, pp. 371-393.

BROWN, T.; CUETO, M.; FEE, E. The World Health Organization and the transition from "international" to "global" public health. *American Journal of Public Health.* 2006, Vol. 96, No. 1, pp. 62-72. DOI: 10.2105/AJPH.2004.050831.

BRUNDTLAND, G. Our Common Future [Report of the World Commission on Environment and Development]. Oslo. 1987. http://www.un-documents.net/our-common-future.pdf.

CIOT, M. *Negotiation and Foreign Policy Decision Making*. 1st ed. 2014. Cambridge: Cambridge Scholars Publishing. ISBN 978-1-4438-5661-4.

CONCEIÇÃO, P. Assessing the provision status of global public goods. In KAUL, I.; CONCEIÇÃO, P.; Le GOULVEN, K.; MENDOZA, R. (Eds.) *Providing global public goods: managing globalization*. Oxford: Oxford University Press, 1st ed. 2002, pp. 152-180. ISBN 0-19-515740-0. DOI: DOI:10.1093/0195157400.003.0007.

DE LOMBAERDE, P.; LANGENHOVE, L. Monitoring and evaluating the provision of (donor-funded) regional public goods. *Regions & Cohesion.* 2011, Vol. 1, No. 1, pp. 101-123. DOI: 10.3167/reco.2011.010107.

EICHNER, A.; KREGEL, J. An essay on post-Keynesian theory: a new paradigm in economics. *Journal of Economic Literature*. 1975, Vol. 13, No. 4, pp. 1293-1314.

EINARSSON, A. The Economic Importance of Culture. *Journal of Mal and Menning*. 2001, Vol. 62, No. 3, pp. 43-50.14

ERSOY, B. Globalization and Global Public Goods. In: PACHURA (ed.) *New Knowledge in a New Era of Globalization*. In Tech Open Access Publisher, 1st ed. 2011, pp. 225-242. ISBN 978-953-307-501-3. DOI: 10.5772/21520.

FRIEDEN, J. The Governance of International Finance. *Annual Review of Political Science.* 2016, Vol. 19, No. 2, pp. 33-48. DOI: 10.1146/annurev-polisci-053014-031647.

GOLDTHAU, A. Governing global energy: existing approaches and discourses. *Current Opinion in Environmental Sustainability.* 2011, Vol. 3, No. 4, pp. 213-217. DOI: 10.1016/j.cosust.2011.06.003.

GRAAFF, J. *Theoretical welfare economics.* 2nd ed. 1968. Cambridge: Cambridge University Press. ISBN 978-0-521-09446-7.

GUILLAUMONT, P. Linkages between official development assistance and global public goods. In: KAUL, I.; Le GOULVEN, K; SCHNUPF, M. (Eds.): *Global Public Goods Financing: New Tools for New Challenges Global Public Goods Financing*. New York: UNDP, 1st ed. 2002, pp. 106-111. ISBN 978-9211261448.

GUPTA, J.; IVANOVA, A. Global energy efficiency governance in the context of climate politics. *Energy Efficiency*. 2009, Vol. 2, No. 4, pp. 339-352. DOI: 10.1007/s12053-008-9036-4.

HAGE, J. Separating rules from normativity. In: ARASZKIEWICZ, M.; BANAS, P.; GIZBERT-STUDNICKI, T.; PLESZKA, K. (Eds.) *Problems of normativity, rules and rule-following.* Springer International Publishing, 1st ed. 2015, pp. 13-29. ISBN 978-3-319-09374-1. DOI: 10.1007/978-3-319-09375-8_2.

HARBERGER, A. Three basic postulates for applied welfare economics: An interpretive essay. *Journal of Economic literature*. 1971, Vol. 9, No. 3, pp. 785-797.

HEAD, J. Public goods and public policy. *Public Finance.* 1962, Vol. 17, No. 1, pp. 197-221.

HEAD, J. Public Goods: The Polar Case Reconsidered. *Economic Record.* 1977, Vol. 53, No. 2, pp. 227-238.

HENDERSON, K. Alternative service delivery in developing countries: NGOs and other non-profits in urban areas. *Public Organization Review.* 2002, Vol. 2, No. 2, pp.99-116. DOI: 10.1023/A:1016051211179.

HERTZFELD, H.; WILLIAMSON, R. The Social and Economic Impact of Earth Observing Satellites. In: Dick, S.; Launius, R. (Eds.) *Societal Impact of Spaceflight.* Washington D.C.: NASA, 1st ed. 2007, pp. 237-266. ISBN 978-0160801907.

HEUMESSER, C.; FRITZ, S.; OBERSTEINER, M.; PEARLMAN, J.; KHALSA, Benefits and challenges of voluntary contribution to GEOSS. *Space Policy.* 2012, Vol. 28, No. 4, pp. 244-252. DOI: doi:10.1016/j.spacepol.2012.09.011.

CHEN, L.; EVANS, T.; CASH, R. Health as a global public good. In INGE K.; GRUNBERGR, I.; STERN M. (Eds.) *Global Public Goods: International Cooperation in the 21st Century Global public goods.* Oxford: Oxford University Press, 1st ed. 1999, pp. 284-304. ISBN 0-19-513051-0. DOI: 10.1093/0195130529.001.0001.

KADDOUS, C. Introduction: the European Union in International Organizations – Functional Necessity or General Aspiration? In: KADDOUS, C. (ed.) *The European Union in International Organisations and Global Governance: Recent Developements.* Oxford: Hart Publishing. 1st ed. 2015., pp. 1-24. ISBN 978-1849467001.

KARLSSON-VINKHUYZEN, S.; JOLLANDS, N. a STAUDT, L. Global governance for sustainable energy: The contribution of a global public goods approach. *Ecological Economics.* 2012, Vol. 83, pp. 11-18. DOI: DOI: 10.1016/j.ecolecon.2012.08.009.

KAUL, I. Global public goods and the poor. *Development.* 2001b, Vol. 44, No. 1, pp. 77-84. DOI: 10.1057/palgrave.development.1110218.

KAUL, I. Global public goods: A key to achieving the millennium development goals [ODS Discussion Draft]. Paris. 2005. http://www.globusetlocus.org/ImagePub.aspx?id=75940.

KAUL, I. Global public goods: what role for civil society? *Nonprofit and Voluntary Sector Quarterly.* 2001a, Vol. 30, No. 3, pp. 588-602. DOI: 10.1177/0899764001303013.

KAUL, I.; CONCEIÇÃO, P.; LE GOULVEN, K.; MENDOZA, R. Overview. In KAUL, I.; CONCEIÇÃO, P.; Le GOULVEN, K.; MENDOZA, R. (Eds.) *Providing global public goods: managing globalization*. Oxford: Oxford University Press, 1st ed. 2002, pp. 1-58. ISBN 0-19-515740-0. DOI: 10.1093/0195157400.003.0001.

KAUL, I.; GRUNBERG, I.; STERN, M. Defining Global Public Goods. In INGE K.; GRUNBERGR, I.; STERN M. (Eds.) *Global Public Goods: International Cooperation in the 21st Century Global public goods.* Oxford: Oxford University Press, 1st ed. 1999a, pp. 2-19. ISBN 0-19-513051-0.

KAUL, I.; GRUNBERG, I.; STERN, M. Global public goods: concepts, policies and strategies. In INGE K.; GRUNBERGR, I.; STERN M. (Eds.) *Global Public Goods: International Cooperation in the 21st Century Global public goods.* Oxford: Oxford University Press, 1st ed. 1999b, pp. 450-507. ISBN 0-19-513051-0. DOI: 10.1177/0899764001303013.

KAUL, I.; MENDOZA, R. Advancing the Concept of Public Goods. In KAUL, I.; CONCEIÇÃO, P.; Le GOULVEN, K.; MENDOZA, R. (Eds.) *Providing global public goods: managing globalization*. Oxford: Oxford University Press, 1st ed. 2002, pp. 78-111. ISBN 0-19-515740-0. DOI: 10.1177/0899764001303013.

KEYNES, J. *The General Theory of Employment, Interest, and Money*. 1st ed. 1936. London: Palgrave Macmillan. ISBN 978-0-230-00476-4.

KRASNER, S. D. (1982). Structural causes and regime consequences. *International Organization*, 1982, Vol. *36*, No. 2, pp. 185-205.

LINDAHL, E. Just taxation − a positive solution. In: MUSGRAVE, R.; PEACOCK, A. (Eds.) *Classics in the Theory of Public Finance*. London: Palgrave Macmillan UK, 1st ed. 1958, pp. 168-176. ISBN: 978-1-349-23428-8.

LOUAFI, S. Global Public Good. In: MORIN, J.; ORSINI A. (Eds.) *Essential Concepts of Global Environmental Governance*. London: Routledge, 1st ed. 2014, pp. 84-85. ISBN 978-0415822473.

MANKIW, G.; ROMER, D. *New Keynesian Economics: Coordination failures and real rigidities vol. 2.* 1st ed. 1991. Massachusetts: MIT Press. ISBN 978-0262631341.

MASCARENHAS, R.; SANDLER, T. Donors' Mechanisms for Financing International and National Public Goods: Loans or Grants? *The World Economy.* 2005, Vol. 28, No. 8, pp. 1095-1117. DOI: 10.1111/j.1467-9701.2005.00721.x.

MAZZOLA, U. The formation of the prices of public goods. In: MUSGRAVE, R.; PEACOCK, A. (Eds.) *Classics in the Theory of Public Finance*. London: Palgrave Macmillan UK, 1st ed. 1958, pp. 37-47. ISBN 978-1-349-23428-8.

MILINSKI, M.; SEMMANN, D.; KRAMBECK, H.; MAROTZKE, J. Stabilizing the Earth's climate is not a losing game: Supporting evidence from public goods experiments. *Proceedings of the National Academy of Sciences of the United States of America*. 2006, Vol. 103, No. 11, pp. 3994-3998.

MOON, S. Medicines as global public goods: The governance of technological innovation in the new era of global health. *Global Health Governance*. 2009, Vol. 2, No. 2, pp. 1-23.

MORRISSEY, O.; TE VELDE, D.; HEWITT, A. Financing international public goods. In FERRONI, M.; MODY, A. (Eds.) *International Public Goods. Incentives, Measurement and Financing*. New York: World Bank Publications. 1st ed. 2002, pp. 119-156. ISBN 0-8213-5110-9.

MUSGRAVE, R. *The Theory of Public Finance: A Study in Public Economy*. 1st ed. 1959. New York: McGraw-Hill. ISBN 978-0070441156.

MUSGRAVE, R. The voluntary exchange theory of public economy. *The Quarterly Journal of Economics*. 1939, Vol. 53, No. 2, pp. 213-237.

NORDHAUS W. Global public goods and the problem of global warming. [The Institut d'Economie Industrielle]. Paris, 1999. http://idei.fr/sites/default/files/medias/doc/conf/annual/paper_1999.pdf.

OLIVER, M.; WILLEM TE VELDE, D.; HEWITT, A. Defining International Public Goods: Conceptual Issues. In FERRONI, M.; Mody, A. (Eds.) *International Public Goods. Incentives, Measurement and Financing*. New York: World Bank Publications. 1st ed. 2002, pp. 31-46. ISBN 0-8213-5110-9.

OLSON, M.; ZECKHAUSER, R. An economic theory of alliances. *The Review of Economics and Statistics*. 1966, Vol. 48, No. 3, pp. 266-279.

PERSSON, A. Aid. In: MORIN, J.; ORSINI A. (Eds.) *Essential Concepts of Global Environmental Governance*. London: Routledge, 1st ed. 2014, pp. 1-3. ISBN 978-0415822473.

PIGOU, A. *The economics of welfare*. 4th ed. 1932. London: Palgrave Macmillan. ISBN 978-1596059504.

PINHEIRO, F.; VASCONCELOS, V.; SANTOS F.; PACHECO J. Evolution of All-or-None strategies in repeated public goods dilemmas. *PLOS Comput Biol*. 2014, Vol. 10, No. 11, pp. 1-5. DOI: 10.1371/journal.pcbi.1003945.

PULKOWSKI, D. (2014). *The law and politics of international regime conflict*. 1.vyd. 2014. Oxford: Oxford University Press, ISBN 9780199689330. DOI: 10.1093/acprof:oso/9780199689330.001.0001.

RAFFER, K. ODA and global public goods: A trend analysis of past and present spending patterns [ODS Background Paper]. New York. 1999. http://homepage.univie.ac.at/kunibert.Raffer/UNDP.pdf.

REILLON, V. *Horizon 2020 budget and implementation: A guide to the structure of the programme.* Brussels: 1st ed. 2015. ISBN 978-92-823-8317-9. DOI: 10.2861/40805.

ROBINSON, J.; EATWELL, J. *An introduction to modern economics*. 1st ed. 1973. New York: McGraw-Hill. ISBN 978-0070840256.

SAMUELSON, P. Aspects of public expenditure theories. *The Review of Economics and Statistics*. 1958, Vol. 40, No. 4, pp. 332-338.

SAMUELSON, P. The pure theory of public expenditure. *The review of economics and statistics*, 1954, Vol. 36, No. 4, pp. 387-389.

SANDLER, T. Assessing the optimal provision of public goods: In search of the holy grail. In KAUL, I.; CONCEIÇÃO, P.; Le GOULVEN, K.; MENDOZA, R. (Eds.) *Providing global public goods: managing globalization*. Oxford: Oxford University Press, 1st ed. 2002a, pp. 131-151. ISBN 0-19-515740-0. DOI: 10.1093/0195157400.003.0006.

SANDLER, T. *Collective action: Theory and applications*. 1st ed. 1992. Ann Arbor: University of Michigan Press. ISBN 978-0472065011.

SANDLER, T. Financing international public goods. In FERRONI, M.; Mody, A. (Eds.) *International Public Goods. Incentives, Measurement and Financing*. New York: World Bank Publications. 1st ed. 2002b, pp. 81-117. ISBN 0-8213-5110-9.

SANDLER, T. Global and regional public goods: a prognosis for collective action. *Fiscal studies*. 1998, Vol. 19, No. 3, pp. 221-247.

SAURIN, J. Global environmental crisis as the 'disaster triumphant': The private capture of public goods. *Environmental Politics*. 2001, Vol. 10, No. 4, pp. 63-84. DOI: 10.1080/714000578.

SAX, E. The valuation theory of taxation. In: MUSGRAVE, R.; PEACOCK, A. (Eds.) *Classics in the Theory of Public Finance*. London: Palgrave Macmillan UK, 1st ed. 1958, pp. 177-189. ISBN: 978-1-349-23428-8.

SEN, A. Distribution, transitivity and Little's welfare criteria. *The Economic Journal*. 1963, Vol. 73, No. 292, pp.771-778.

SHACKLE, G. L. S. *Expectation Enterprise and Profit: The Theory of the Firm.* 1970. London: Geogre Allien and Unwin Ltd. ISBN 0/04/330160/6.

SHEPSLE, K.; WEINGAST, B. The Positive Theory of Public Goods Provision. A paper presented at the 21st Annual Conference of the European Association of Environmental and Resource Economists, Helsinki, Finland, 2 April 2014, pp.1-42.

SCHAPIRO, C.; STIGLITZ, J. Equilibrium unemployment as a worker discipline device. *The American Economic Review*, 1984, Vol. 74, No. 3, pp.433-444.

SCHMIDT, S.; WONKA A. European Commission. In: JONES, E.; MENON, A.; WEATHERILL, S. *The Oxford Handbook of the European Union*. Oxford: Oxford University Press. 1st ed. 2012, pp. 336-349. ISBN 978-0199546282. DOI: 10.1093/oxfordhb/9780199546282.013.0024.

SCHULZE-GÄVERNITZ, G. *Volkswirtschaftliche Studien aus Russland*. 1899. Duncker & Humblot.

SMITH, R.; BEAGLEHOLE, R.; WOODWARD, D.; DRAGER, N.; SMITH, R. *Global public goods for health: health economic and public health perspectives.* 1st ed. 2003. Oxford: Oxford University Press. ISBN 978-0198527985. DOI: 10.1002/hec.905.

SONNTAG, D. Funding HIV-Vaccine Research In Developing Countries—What Is Wrong With Iavi's Recommendation. *Health Economics*, Vol. 23, No. 2, pp. 141-158. DOI: 10.1002/hec.2909.

STIGLITZ, J. The Overselling of Globalization. In: WEINSTEIN, M. (ed.) *Globalization: What's New?* New York: Columbia University Press. 2nd ed. 2013, pp. 228-262. ISBN 978-023-113-4590.

TAVOR, T.; SPIEGEL, U. The optimal supply of congested public goods for homogeneous and heterogeneous customers. *The Journal of International Trade & Economic Development.* 2016, Vol. 25, No. 1, pp. 103-130. DOI: 10.1080/09638199.2015.1040054.

This is the translated version of the paper that was already published in Czech language. Please Please cite as follows: Machoň, Miloslav (2017). Global Public Goods: The Case for the Global Earth Observation System of Systems. Acta Oeconomica Pragensia, 25(3), 68-83. https://doi.org/10.18267/j.aop.583TURNER, B.; HOLTON, R. Theories of globalization: issues and origins. In: TURNER, B.; HOLTON, R. (Eds.) *The Routledge International Handbook of Globalization Studies*. Routledge, 2nd ed. 2015, pp. 3-24. ISBN 978-1317964919. DOI: 10.4324/9781315867847.

UNNEVEHR, L. Mad cows and Bt potatoes: global public goods in the food system. *American Journal of Agricultural Economics*. 2004, Vol. 86, No. 5, pp. 1159-1166. DOI: 10.1111/j.0002-9092.2004.00661.x.

VOGLER, J. Studying the global commons: governance without politics? *Handbook of Global Environmental Politics.* Cheltenham: Edward Elgar Publishing, 3. vyd. 2012, pp. 172-183, ISBN 978-1781005446. DOI: 10.4337/9781849809405.00023.

WEIR, L.; MYKHALOVSKIY, E. The geopolitics of global public health surveillance in the twenty-first century. In: **Bashford, A. (Ed.)** *Medicine at the Border*. London: Palgrave Macmillan UK, 1st ed. 2014, pp. 240-263. ISBN 978-0-230-28890-4. DOI: 10.1057/9780230288904_13.

## Sources

EC 1992. Treaty on European Union. Maastricht, 7. 2. 1992. http://eur-lex.europa.eu/legal-content/EN/TXT/PDF/?uri=OJ:C:1992:191:FULL&from=EN.

EC 2007. Treaty of Lisbon: Amending the Treaty on European Union and the Treaty Establishing the European Community, Lisabon, 17. 12. 2007. http://publications.europa.eu/resource/cellar/688a7a98-3110-4ffe-a6b3-8972d8445325.0007.01/DOC_19.

EC 2012. Consolidated version of the Treaty on the Functioning of the European Union. Brussels, 26. 10. 2012. http://eur-lex.europa.eu/legal-content/EN/TXT/PDF/?uri=CELEX:12012E/TXT&from=EN.

EC 2014. Commission Staff Working Document Global Earth Observation System of Systems (GEOSS): Achievements to date and challenges to 2025. Brussels, 25. 9. 2014. http://ec.europa.eu/transparency/regdoc/rep/10102/2014/EN/10102-2014-292-EN-F1-1.Pdf.

EC 2015. The EU's Framework Programme for Research and Technological Development is a major tool to support the creation of the European Research Area – ERA. Brussels, 11. 08. 2015. https://ec.europa.eu/research/fp6/index_en.cfm.

G8 2003. Science and Technology for Sustainable Development: A G8 Action Plan. Toronto: G8. http://www.g8.utoronto.ca/summit/2003evian/sustainable_development_en.html.

GEO 2005a. Ad Hoc GEO documents. Geneva: GEOSS. http://www.earthobservations.org/documents.php?smid=500.

GEO 2005b. Resolution of the Third Earth Observation Summit. Geneva: GEOSS. https://www.earthobservations.org/documents/eos_iii/Third%20Summit%20Resolution.pdf.

GEO 2005c. The Global Earth Observation System of Systems (GEOSS) 10-Year Implementation Plan. Geneva: GEOSS. http://www.earthobservations.org/documents/10-Year%20Implementation%20Plan.pdf.
19

GEO 2006. GEO Resources and Expenditure for 2006. Geneva: GEOSS.

GEO 2007. GEO Interim Report on Income and Expenditure 2013. Geneva: GEOSS.

GEO 2008. GEO Report on Income and Expenditure 2008. Geneva: GEOSS.

GEO 2009. GEO Report on Income and Expenditure 2009. Geneva: GEOSS.

GEO 2010. GEO Interim Report on Income and Expenditure 2010. Geneva: GEOSS.

GEO 2011. GEO Interim Report on Income and Expenditure 2011. Geneva: GEOSS.

GEO 2012. GEO Interim Report on Income and Expenditure 2012. Geneva: GEOSS.

GEO 2013. GEO Interim Report on Income and Expenditure 2013. Geneva: GEOSS.

GEO 2014a. GEO: Rules of Procedure. Geneva: GEOSS. https://www.earthobservations.org/documents/GEO%20Rules%20of%20Procedure.pdf.

GEO 2014b. GEO Interim Report on Income and Expenditure 2014. Geneva: GEOSS.

GEO 2015a. The GEO Plenary. Geneva: GEOSS. http://www.earthobservations.org/governance.php.

GEO 2015b. GEOSS Portal: Copyright. Geneva: GEOSS. http://www.earthobservations.org/g_copyright.html.

GEO 2015c. GEOSS Portal - Terms of Use. Geneva: GEOSS. http://www.geoportal.org/web/guest/geo_terms_and_conditions.

GEO 2015d. GEO: Member List. Geneva: GEOSS. http://www.earthobservations.org/members.php.

GEO 2015e. GEOSS Portal. Geneva: GEOSS. http://www.geoportal.org/web/guest/geo_home_stp.

GEO 2015f. GEO Interim Report on Income and Expenditure 2015. Geneva: GEOSS.

UN 2015. Member States of the United Nations. Geneva: GEOSS. http://www.un.org/en/members/.

WMO 2003. Issues Relevant to CBS Arising From Fourteenth Congress, the Fifty-Fifth Session of The Executive Council, and the 2003 Meeting of the Presidents of Technical Commissions Earth Observation Summit. Geneva: WMO. http://www.wmo.int/pages/prog/www/BAS/MG-4/Doc2(7).pdf.

I acknowledge Ania Pavlickova Wilinska, MA for translating the paper to the English language.